\documentclass{llncs}

\input{psfig.sty}

\begin{document}

\pagestyle{empty}

\mainmatter

\title{C++ programming language for an abstract massively parallel SIMD architecture}

\titlerunning{SIMD C++}

\author{Alessandro Lonardo\inst{1} \and Emanuele Panizzi\inst{2,1}
\and Benedetto Proietti\inst{3}}


\institute{Istituto Nazionale di Fisica Nucleare (INFN),
P.le Aldo Moro, 2, 00185 Roma, Italy\\
\and Universit\`a degli Studi dell'Aquila, Dipartimento di
Ingegneria Elettrica,\\
Monteluco di Roio, 67100 L'Aquila, Italy\\
\and Universit\`a degli Studi ``La Sapienza'', Dipartimento di Matematica\\
P.le Aldo Moro, 2, 00185 Roma, Italy\\
\email{\{alessandro.lonardo, emanuele.panizzi,
benedetto.proietti\}@roma1.infn.it}}

\maketitle

\begin{abstract}
The aim of this work is to define and implement an extended C++
language to support the SIMD programming paradigm.
The C++ programming language has been extended to express all the
potentiality of an abstract SIMD machine consisting of a central
Control Processor and a N-dimensional toroidal array of Numeric
Processors.
Very few extensions have been added to the standard C++ with the goal
of minimising the effort for the programmer in learning a new language
and to keep very high the performance of the compiled code.
The proposed language has been implemented as a porting of the GNU C++
Compiler on a SIMD supercomputer.
\end{abstract}

\section{Introduction}

The aim of this work is to define and implement an extended C++
\cite{strstr3}
language to support the SIMD \cite{hwangbriggs} programming paradigm.
Our goal is to add minimal extensions to the
standard C++ language \cite{ansidraft} in order to minimise the
syntactical differences
when porting standard C++ applications or writing new codes. Our
decision to be always as close as possible to the standard lead to
the definition of an extended C++ language with very few constructs to learn for C++
programmers, and relatively
easy to use.

Using our language, the SIMD parallelism is efficiently achieved with
traditional sequential programming plus a couple of new constructs
(used to perform memory mapped internode communication and to inhibit
execution of code on some processing nodes)
and some knowledge of the native data types and their allocation.
The programmer can thus focus on the realization of the algorithm
and on the data distribution, which are the key points to exploit the
parallel architecture.

The proposed language has been implemented as a porting of the GNU C++
Compiler \footnote{Release 2.95.1} \cite{PortingGnuCC} for the APEmille parallel supercomputer
\cite{APE:chep,APE:parcomp}. Some modifications of the GNU C++ Compiler
have
been introduced, as well as the complete re-definition of the back-end
for the target machine \cite{PortingGnuCC}. APEmille is a parallel
SIMD computer developed at INFN (Italian National Institute for
Nuclear Physics) capable of peak performance of 1 Teraflop in a
configuration with 2048 processing nodes.

The simplicity and low number of extensions to the standard language
helped reaching the goal of
efficiency of the executable parallel codes, main goal
for any number crunching application running on a massively parallel
supercomputer.

In this paper, we describe the proposed SIMD C++ language, and
especially those aspects which extend the standard C++ syntax or
semantics. Section \ref{sect:SIMDC++} is devoted to this
description. Section \ref{sect:abstractSIMD} explains the abstract
SIMD architecture which we refer to, while section
\ref{sect:relatedwork} reports on those works related to our either for
the language used (extensions of C/C++) or for a similar target
architecture or parallel paradigm. Section \ref{sect:Concl} contains the
conclusions.

\section{The Abstract SIMD machine}
\label{sect:abstractSIMD}

SIMD machines consists of synchronized processing elements with an
associated unique control processor. The Control Processor (later, CP)
broadcasts the same
instruction stream to all processing elements. All processing elements
execute the same instruction at each clock
cycle on their own data.
In the proposed architecture the processing elements are specialized
processors
for numeric applications: we will call them Numeric Processors or NPs.
The NPs form an N-dimensional toroidal array.
Each NP consists of an ALU, an own register file, a local memory and a
local
memory mass storage.
There is no shared memory across the whole machine: communication
among NPs is achieved through a memory mapping mechanism that allows
each NP to access memory of its neighbours.
The machine is SIMD and guarantees no conflicts in memory accesses.

\subsection{Control Processor}
The CP handles integer data types, executes branches, function calls,
and
generates memory address.
Every instruction is sent by the CP to each NP in the machine, at the
same clock cycle.

Global addresses are broadcasted to all NPs. They will use them to
address their own memory.

\subsection{Numeric Processor}

NPs are specialized in numeric instructions, in fact they natively
support
floating point data types -- scalar (both single and double precision),
vector/complex (couple of single precision) and the integer data type.

NPs, in order to perform conditional execution, can test local
conditions, and, when they are not met, can disable the effect of the
following numeric instructions.
We call the conditional test a {\it where} instruction.

CP can also make all NPs test their local conditions. Then it can  perform a
global branch
if {\it any}, {\it all} or {\it none} of the NPs has met the condition.

NPs can address their own memory using the global address generated by
the CP and, eventually, adding a local offset.

\section{Proposed SIMD C++ language}
\label{sect:SIMDC++}

In this section we describe our extended C++ analyzing all the aspects
more related to parallelism. Subsection \ref{subsect:remarks} contains
a resume of the main characteristics and discusses general topics.
\subsection{Types, Declarations and Allocation}
\subsubsection{Basic Data Types}

With {\it basic data types} or {\it basic types} we refer to the types
natively
supported by the language, as \verb|int| or \verb|float|.
The C++ language proposed in this paper includes all the basic
data types supported by the illustrated abstract SIMD machine.
Namely:
\begin{verbatim}
int, float, double, complex, vector, localint
\end{verbatim}
The \verb|localint| data type are integer variables allocated in the numeric processors (see
later), while the types \verb|vector| and \verb|complex| represent a
pair of float and are treated as native types by our abstract machine.
Pointers, arrays and function pointers are supported for every type and
every level of indirection.

These types are all signed. All the "standard" C++ types (e.g.: \verb|long long|, \verb|long double|, \verb|char|, etc. and all
unsigned types) could be supported, performing
software emulation for those not supported by the physical machine.

Declaring variables is absolutely identical to standard C++.
It is also possible to declare new types with the \verb|typedef|
keyword just as in C++, with the standard rules and no limitation.

\subsubsection{Allocation}

The variables declared are allocated in the Control Processor (CP) or
in the Numeric Processor (NP) depending on their type.
We divide the basic types into two groups: the Control Processor types
(\verb|int|, pointers) and the Numeric Processor types (\verb|float|, \verb|double|, \verb|vector|,
\verb|complex|, \verb|localint|).

CP variables are allocated in the (unique) Control Processor,
so there is just one instance of them; NP variables are allocated
in {\it each} NP data memory, and, most important, at the same location
in
 every one: so there are multiple instances of numeric variables.
This is the essence of the SIMD programming paradigm, and implies a
very important fact: {\it memory images of NPs are all
identical; each allocation, both static and dynamic, is the same for
every NP.}

Arrays of any type are allocated in the same processor of the base
type. For example, the array
\begin{verbatim}
double a[100000];
\end{verbatim}
is allocated in the NPs' memories.
However the base of the array, which is known at compile time,
is a pointer, and so it is handled by the CP.

The allocation mechanism is automatic and controlled by the compiler:
there is
no way and no reason for the programmer to alter it. On the other hand
data distribution is left to the programmer. We will discuss this
topic in subsection \ref{subsect:imple}.

\subsection{Expressions}
\subsubsection{Expressions within the same type}

Handling expressions among variables of the same type is not ambiguous,
because
they are allocated in the same kind of processor so code for "that"
processor
will be generated to handle them. For example:

\vbox{
\noindent
\fbox{\it CP allocation}\verb|   int i,j,k;| \\
\verb|                double a,b,c;   | \fbox{\it NP allocation}\\
\fbox{\it CP code} \verb|       i = j+1;| \\
\verb|                a = 1.0;        | \fbox{\it NP code}\\
\verb|                b = a*c-b;      | \fbox{\it NP code}\\
\fbox{\it CP code} \verb|       k++;|  }

\begin{table}[htpp]
\begin{center}
\begin{tabular}{|l||c|c|c|c|c|c|c|c|}
\hline
Promotions $>>$ & \verb|int| & CP ptr & NP ptr &\verb|float|& \verb|double|
&
\verb|vector|& \verb|complex| & \verb|localint| \\
&&&&&&&&\\
\hline
\hline
\verb|int| & yes & yes & yes & yes & yes & yes & yes & yes \\
\hline
CP pointer & yes & yes & yes & no & no & no & no & no \\
\hline
NP pointer & yes & yes & yes & no & no & no & no & no \\
\hline
\verb|float| & no & no & no & yes & yes & yes & yes & yes \\
\hline
\verb|double| & no & no & no & yes & yes & yes & yes & yes \\
\hline
\verb|vector| & no & no & no & no & no & yes & no & no \\
\hline
\verb|complex| & no & no & no & no & no & no & yes & no \\
\hline
\verb|localint| & no & no & yes & yes & yes & yes & yes & yes \\
\hline
\noalign{\vskip 0.5cm}
\end{tabular}
\caption{Allowed Promotions}
\end{center}
\end{table}

\subsubsection{Mixed-types expressions}
They are handled by promoting types, or by explicit cast by the
programmer.
There are specific rules about cast and promotions.\\
\begin{itemize}
\item cast/promotions from CP to NP types are ALWAYS allowed.
\item cast/promotions from NP to CP types are NEVER allowed.
\item cast/promotions between two types of the same group are
allowed depending on the specific types.
\end{itemize}

\begin{table}[htpp]
\begin{center}
\begin{tabular}{|c|c|c|c|c|c|c|c|c|}
\noalign{\vskip 0.5cm}
\hline
Casts $>>$ & \verb|int| & CP ptr & NP ptr &\verb|float|& \verb|double| &
\verb|vector|& \verb|complex| & \verb|localint| \\
\hline
\verb|int| & yes & no & no & yes & yes & yes & yes & yes \\
\hline
CP pointer & no & yes & no &no & no & no & no & no \\
\hline
NP pointer & no & yes & no & no & no & no & no & no \\
\hline
\verb|float| & no & no & no & yes & yes & yes & yes & no \\
\hline
\verb|double| & no & no & no & no & yes & no & no & no \\
\hline
\verb|vector| & no & no & no & no & no & yes & no & no \\
\hline
\verb|complex| & no & no & no & no & no & no & yes & no \\
\hline
\verb|localint| & no & no & no & no & no & no & no & yes \\
\hline
\noalign{\vskip 0.5cm}
\end{tabular}
\caption{Allowed Casts}
\end{center}
\end{table}

It is obvious that a cast/promotion from a CP type to NP generates
multiple
instances of one value.

\subsection{Multiple Addressing}

As stated before, the abstract SIMD machine includes the ability to add
a local
offset when accessing local NP memory, so every NP could access a
different location in memory.
This is realized with the \verb|localint| variables, that are integers
 allocated in the NPs.
These values can be used to add a local displacement when accessing
local memory.

A pseudo function \verb|localoffset()| can be called with a \verb|localint|
argument to set the local offset for the following memory access.

\begin{verbatim}
int i;
localint li;
float r, a[100];
// ...
localoffset(li);
r = a[i];
\end{verbatim}

In the example above, access in array \verb|a| is at index $i + li$.

\subsection{Type Constructors: Structs, Classes and Unions}
\label{sect:record_types}
It is possible to declare a new type using \verb|struct|,
\verb|class| or \verb|union| as in standard C++.
Structs and classes can contain data fields of any other data type
(both CP types and NP types),
while unions must contain fields associated to the same kind of
processor (only CP or only NP), due to allocation reasons, as explained
before.

\begin{verbatim}
class Mixed {
   int a;
   float x;
public:
   Mixed (int aa, float xx) : a(aa), x(xx) {};
};
\end{verbatim}

Each field is allocated in the respective processor so that multiple
instances
of numeric field exist. The effort to address them and to keep
pointers consistent is made by the compiler.
Fields must be accessed directly with pointers: incrementing and
decrementing
pointers to "navigate" through a struct or class could generate
unpredictable results because the object is allocated on different
memories.
The space allocated is compacted, so that only the necessary size is
allocated in each kind of processor.

\subsection{Object Oriented Features: \\Encapsulation, Inheritance,
Polymorphism}

Encapsulation is handled as in standard C++ with no other extension nor limitation.
Field allocation follows what stated in \ref{sect:record_types}.
Methods are called passing them the invocation object as an hidden
argument.The same method is executed by each NP.

Also Inheritance and Polymorphism have no extensions nor limitations.
Non-virtual base class members are inserted in the CP or NP instance
layout of the object after their type class.
Virtual base class members and virtual classes information are inserted
into
the CP instance of the object: in fact they are pointers.

\subsection{Communication}
Communication among Numeric Processors is achieved through memory
mapping. The proposed C++ language allows to address an array element
in a remote NP by summing a constant to the array index or to the
pointer that would be used for local access.
Different pre-defined constants are associated to neighbour NPs. These
constants specify the relative position of the NP to be accessed with
respect to the current NP. The constants are
generated and handled on the CP, so they are the same for all NPs.
The following example shows communication between NPs:
\begin{verbatim}
  float r, v[100000];
  r = v[3+XPLUS_NP];  // each NP accesses the 3rd element of
                      // the nearest neighbour on the x axis
                      // in the positive direction
\end{verbatim}
The constant \verb|XPLUS_NP| is machine dependent.

It is possible to use a remote object as parameter or invocation object of a
method. In this way, a code like:
\begin{verbatim}
class C
{public:
  float x;
  void f(float y) { x = y; }
};

int main()
{ float a;
  C v[10];
  // ...
  v[0+XPLUS_NP].f(a);
  // ...
}
\end{verbatim}
assigns to the \verb|x| field of the \verb|v[0]| object on each node the value of \verb|a| on
the adjacent (\verb|XMINUS_NP|) node.

\subsection{Local Conditions}
The instruction flow being unique, it is possible to branch only when
global
conditions (conditions on the CP) are met.
Conditions on local variables (on NPs) can be handled with the
\verb|where|-\verb|elsewhere|
keywords\footnote{In our implementation, these are not keywords but
function names}. Conditioned code will be executed only
by those NPs that met the condition. All the other NPs will execute
NOPs. CP instructions
inside a where block, on the other hand, are always executed.
\verb|where|-\verb|elsewhere| are used just like \verb|if|-\verb|else|,
as shown in the following example.

\begin{verbatim}
int i;
double x,y;
// ...
where (x != 0.0)
{ y = 1/x;
}
elsewhere
{ y = 0;
}
\end{verbatim}

\subsection{Examples of implementation of SIMD programs}
\label{subsect:imple}
When writing a program for a SIMD machine using the proposed C++
extensions,
the ``SI'' part of the SIMD paradigm is realized using a single
instruction stream, as the C++ language naturally does;
the ``MD'' part of SIMD is achieved allocating multiple instances of
the numeric variables.

The initial loading of different data in each Numeric Processor data
memory is made by the operating system, while the slicing of a big
array into the NPs must be determined by the programmer.

For example, suppose that the problem needs to handle an array of
$10000\times10000$
elements, and that the target machine has a 2D square
topology with $10\times10=100$ Numeric Processors. The programmer will
declare a $1000\times1000$ array (and every NP will have its own
instance of this array, that represents a slice of the big array).
A very trivial example of code that implements the sum of two
$10000\times10000$ elements
arrays can be useful to explain the parallelization mechanism and the
data distribution:
\begin{verbatim}
int main()
{
   const int dimx = 1000;
   const int dimy = 1000;
   const int size_per_node = dimx*dimy;

   float m1[dimx][dimy], m2[dimx][dimy], m3[dimx][dimy];
   const char *filename = "myfile.data";
   // ...
   distributed_load(m1, filename, size_per_node);
   distributed_load(m2, filename, size_per_node);
   for (int i=0; i<dimx; i++)
     for (int j=0; j<dimy; j++)
       m3[i][j] = m1[i][j] + m2[i][j];
   distributed_store(m3, filename, size_per_node);
  // ...
}
\end{verbatim}

Numeric instructions will be executed in
parallel by the Numeric Processors on their local data, while Control
Processor will execute flow control, integer instructions and will
generate memory addresses.
The \verb|distributed_load()| and \verb|distributed_store()| functions
perform machine dependent system calls supposed to load and store data
in the appropriate way.

\subsection{General remarks on the language proposed}
\label{subsect:remarks}
The proposed C++ is very similar to the standard C++, is easy to
learn, to use and to debug, produces highly efficient executable
codes, and can be used in a professional environment.

The most important aspect of the proposed language is that it is a
minimal extension to the C++ standard. This is a key feature as we
want the programmer to concentrate on the application development
rather than paying attention to implementation aspects.

Our language strictly conforms to the machine architectural characteristics in order
to fully exploit the simplification that 
the SIMD synchronous structure and the memory mapped internode
communications introduce in the task of writing the parallel algorithm.

As a result, there is no need to develop multi-threaded programs nor
to use any special communication library.

Finally, object distribution is obtained by the simple allocation
model described above. All objects are replicated on each processing
element and invocation of a method is executed on all NPs (or on the
subset of them satisfying an eventual WHERE condition).

\section{Related work}
\label{sect:relatedwork}

In this section we compare our language to a couple of parallel C/C++
extensions and to High Performance Fortran, which is the standard for data parallel applications.

HPC++ \cite{HPC++,HPC++2} is a set of class libraries and tools that 
extend the C++ language. It also has a set of runtime systems
that are required for remote access. It  refers to a very general
architectural model, so it can be used on a variety of
machine architectures. There are two main execution modes for
programs written in HPC++: 
\begin{enumerate}
\item multi-thread shared memory mode, suitable for
coarse-grained applications with some particular collective
operations for thread synchronization.
\item Single Program Multiple Data (SPMD) mode, in which $n$ copies of
the same program run on the $n$ processing nodes. This mode is
similar to using C/C++ with MPI or PVM. The programmer must manage the
data distribution and the synchronization of processes.
\end{enumerate}
HPC++ has thus a totally different approach compared to our one, and
this approach
is not applicable to our language architecture which is focused on
single threaded programs. 

pC++ \cite{pC++} is a C++ extension that provides a thread-based programming model and a simple way to
encapsulate SPMD code in it, together with a mechanism for data distribution
similar to the one adopted in HPF (see below).  
The key concept of this extension is the {\em collection} of objects.
It is possible to invoke a method on an entire collection or on a part of it. 
The compilation of pC++ code is achieved as a translation into standard C++ by a
preprocessor.
 
The HPF \cite{HPF} programming model has the following key points:
\begin{enumerate}
\item single threaded control;
\item global namespace, low-level of data distribution and remote
communication details hidden to the programmer;
\item loosely synchronous: synchronization of program execution on
different nodes is accomplished only at special points (e.g. the
completion of a loop) and not instruction by instruction;
\item parallel operations: operations on array elements executed at
the same time over all nodes.
\end{enumerate}

HPF extends the Fortran language adding compiler directives, libraries
and new language constructs.

The most relevant compiler directives to our purposes are those
related to data distribution.  This is accomplished in three
steps: in the first step an alignment is defined for arrays, in the
second step aligned data are mapped on a abstract set of processors
and finally this set is mapped onto
physical processors. This is quite different from our approach as we
force the programmer to take care of allocation of large matrices as
described above. In particular we rely on operating system calls to perform
something similar to the  BLOCK and DEGENERATE distribution types.

A similarity between HPF and our language is the specification of
locally conditioned code execution via the WHERE statement, although
the HPF version accepts logical-array arguments while ours accepts any
logical condition between NP data.
Other parallel constructs, such as FORALL, are not provided by our
language.

HPF is the parallel extension to a standard language that, compared
with the other two, best matches with our approach.

\section {Conclusions}
\label{sect:Concl}
The next topics to be analyzed will include exception
handling, RTTI (Run Time Type Information) and namespaces.

Our implementation of the compiler based on a porting of the GNU CC
compiler on the APEmille architecture is currently under test. We plan
to discuss our implementation in a further paper.

\end{document}